\documentclass[prc,twocolumn,showpacs]{revtex4}
\usepackage{graphicx}

\begin{document}

\title{Tests of the random phase approximation for transition strengths}

\author{Ionel Stetcu}
\altaffiliation{On leave from National Institute for Physics
and Nuclear Engineering Horia Hulubei, Bucharest, Romania}
\affiliation{
Department of Physics and Astronomy,
Louisiana State University,
Baton Rouge, Louisiana 70803-4001}
\author{Calvin W.~Johnson}
\affiliation{Physics Department, 
San Diego State University,
5500 Campanile Drive, San Diego, California 92182-1233}

\begin{abstract}
We investigate the reliability of transition 
strengths computed in the random-phase approximation 
(RPA), comparing with exact results from 
diagonalization in full $0\hbar\omega$ shell-model spaces. The 
RPA and shell-model results are in reasonable agreement 
for most transitions; however some very low-lying collective 
transitions, such as isoscalar quadrupole, are 
in serious disagreement. We suggest the failure 
lies with incomplete restoration of broken symmetries 
in the RPA. Furthermore we prove, analytically and numerically,
that standard statements regarding the energy-weighted sum rule 
in the RPA do not hold if an exact symmetry is broken. 
\end{abstract}
\maketitle

\section{Introduction}

Electromagnetic and weak transitions provide important probes of nuclear
wavefunctions. Exact numerical solutions of the many-body Schr\"odinger
equation are very difficult, so theorists have devised a number of
approximations. Transitions offer a serious test of the model
wavefunctions, not only in comparison to experiment but also between
approximations.

One of the most successful approximations is the interacting shell model (SM), 
which diagonalizes an effective Hamiltonian in a restricted
space, providing good description of the observed low lying states for a
wide range of mass numbers. The SM wavefunctions successfully describe 
electromagnetic and weak transitions, in good agreement (albeit sometimes 
requiring effective couplings) with the
experimental data; for this reason, in this paper 
we consider SM results to be exact. Due to the computational
burden which increases with mass number, however, full $0\hbar\omega$ SM
calculations are presently limited to light and medium nuclei ($A<80$).

Another approximation, which works surprisingly well considering its
simplicity, is mean-field theory, e.g., Hartree-Fock (HF) theory. In
this approach, the particles are uncorrelated and move in an average
potential which replaces the two-body interaction; the ground state is
determined by energy minimization. Starting from the HF solution, the
Tamm-Dancoff approximation (TDA) constructs excited states by assuming
them to be mixtures of one particle, one hole (1p-1h) configurations. The
HF+TDA is the simplest model for ground and excited state wavefunctions.
There are also limitations of this model: the ground state wavefunction is
a single Slater determinant which can break symmetries of the Hamiltonian
and neglects particle-hole correlations arising from the residual
interaction \cite{ring}.

The random phase approximation (RPA) is a generalization of the TDA, which
takes into account the residual interaction by adding 2p-2h correlations
in the ground state.  (An important extension is
 the quasiparticle RPA  or QRPA; because it accounts for 
nonperturbative pairing effects, it is often 
considered a superior approach. In this paper we 
restrict ourselves to number-conserving RPA.) The excited states, as in the TDA,
remain 1p-1h mixtures. Is this the case? That is, are the excited states
\textit{mainly} mixtures of 1p-1h correlations? The answer is no: one
knows from SM calculations that, for example, the first excited state in
$^{16}$O has a predominant 4p-4h character \cite{haxton1990}, outside the
RPA model space. Is therefore no surprise that RPA fails in these
situations.

Despite known limitations, the RPA (or QRPA) has been widely applied for
decades to evaluate spectra and transition strengths in nuclei. The
reason, aside from computational simplicity, is that it is frequently,
although not always \cite{kuo1970,m2strength}, in close agreement with 
experiment. Thus, RPA was the method of choice for description of negative
parity states in closed shell nuclei
\cite{kuo1970,ullah69,blomqvist1969,true1971} and in open shell 
nuclei \cite{rowe1970}. RPA and QRPA calculations
(sometimes including continuum states) using phenomenological interactions have
been successful in describing the experimental position of
giant resonances \cite{Goeke82}, particularly 
E1 \cite{rowe1970,e1} and M1 \cite{soloviev1999} from
electron or proton scattering, or Gamow-Teller resonances
\cite{grimes1996}. In general, however, the description of low-lying
transitions is poor \cite{grimes1996,lowstates}.  Other studies have used
QRPA to compute transitions of interest for astrophysics but did not
directly compare to experiment \cite{jameel,hamamoto}.

What about tests of RPA against exact models? Until recently there was no
thorough test of the HF+RPA against an exact non-trivial model for binding
energies. We performed such a test for a large number of nuclei and found that
in general HF+RPA is a good approximation to the exact shell-model binding
energy \cite{stetcu2002}, although with some significant failures. Tests of 
transitions have been also mainly against  `toy' models 
\cite{lipkinTest1,lipkinTest2,engel1997}, although the QRPA has been tested against exact
diagonalization in the full SM space for $\beta^{\pm}$ or double $\beta$ decays
\cite{lauritzen,civitarese,Zhao1993,auerbach}, with mixed success. 
Broader tests of RPA transitions strengths against an exact model, 
e.g., SM, has not been done. To address this
issue, this paper tests the RPA against full $0\hbar\omega$ calculation
for several nuclei in the \textit{sd} and \textit{pf} shells for
electromagnetic transitions. Unlike previous tests however, we consider
mean-field solutions which break the rotational symmetry.

We find that, in general, the RPA produces a reasonable approximation to
the exact (SM) transitions.  The most significant failures are for
certain low-lying collective states.  We understand these failures through
the incomplete restoration
of broken symmetries and the nature of collective giant resonances.

The Brown-Bosterli schematic model \cite{ring,brown1959} provides insight
into giant collective resonances. It assumes the model Hamiltonian to be
single-particle energies plus a separable residual interaction. In both
the TDA and the RPA, all of the Brown-Bosterli transition strength is to a single state,
the collective state, which is a model for giant resonances.  If the
residual interaction is repulsive, then the collective state will be at high
energy. If the  interaction is attractive, then the collective state will
be low in energy. In more realistic models, of course, the residual
interaction includes more complicated two-body forces, causing the giant
resonance to spread over many states. The important lesson of the
Brown-Bosterli model, however, is that an attractive interaction, such as
isoscalar quadrupole-quadrupole, leads to large collective transitions low
in the spectrum, while repulsive interactions, such as $\sigma \cdot
\sigma$, produce collective transitions lying higher in the spectrum.

Breaking of symmetries can result in low-lying collectivity being subsumed
into the ground state.  For example, the strongly attractive isoscalar
quadrupole-quadrupole interaction leads to a quadrupole deformation in the
HF state.  While the RPA identifies broken symmetries, restoration of
symmetry is incomplete. This point is not fully explicated or understood
in the literature, but our calculations here provide further evidence, as
discussed in detail in Sec.~\ref{sec:upEn}.

The paper is organized as follows. Sec.~\ref{Sec:formalism} presents
briefly the interacting SM and the RPA formalisms, with emphasis on their
application to transition strengths. In Sec.~\ref{Sec:compstr} we compare
the SM and RPA transition strengths and distribution properties for
several nuclides, while in Sec.~\ref{Sec:concl} we summarize results. 
In the Appendix we consider in some detail the effect of breaking of 
exact symmetries on transition strength sum rules.

\section{formalism}
\label{Sec:formalism}

\subsection{General Overview}

In this section we briefly present the treatment of transitions in both SM and
RPA. Before giving any specific details, we review
some general concepts.

A primary goal is, given the Hamiltonian $H$, to calculate solutions to the
eigenvalue equation
\begin{equation}
H|\nu\rangle = E_\nu |\nu \rangle.
\label{eq:eig}
\end{equation}
Suppose we have solved Eq. (\ref{eq:eig}), either in the SM or RPA. The
transition strength from an excited to the ground-state is given by the
square of the matrix element,
\[
S( 0 \rightarrow \nu)=|\langle \nu |F|0\rangle|^2,
\]
while the quantity
\begin{equation}
S_k=\sum_\nu (E_\nu-E_0)^k |\langle \nu |F|0\rangle |^2
\label{srDef}
\end{equation}
is the (energy weighted) sum rule of order $k$. In this equation, $\nu$ runs
over all states, $0$ stands for the ground state, and $F$ is an arbitrary
transition operator. Particularly important is $S_0$, which is
the total transition strength from the ground state to excited states, and 
which can be rewritten as a ground state expectation value:
\begin{equation}
S_0 \equiv  \sum_{\nu}|\langle \nu|F |0\rangle|^2 = \langle 0| F^\dagger F|0\rangle .
 \label{eq:S0gs}
\end{equation}
(In most realistic applications the transition 
operator is a spherical tensor of rank $K$, $F_{KM}$, which 
has the property $F_{KM}^\dagger = (-1)^M F_{K-M}$ \cite{edmonds}. Then 
properly the total strength in Eq.~(\ref{eq:S0gs}) should be 
\begin{equation}
S_0 = \sum_M (-1)^M \langle 0 | F_{K-M} F_{KM} | 0 \rangle,
\end{equation}
and similarly for the double commutator 
in the energy-weighted sum rule, etc.  To avoid clutter we drop the sum over $M$
and it should be assumed to be implicit. )

In fact one can write all the sum rules of order $k$ as expectation values, 
most famously  the linear energy-weighted sum rule:
\begin{equation}
S_1 =\sum_\nu (E_\nu-E_0)
|\langle 0 |F|\nu\rangle|^2 =\frac{1}{2} \langle
0 |[F,[H,F]]| 0\rangle.
\label{genEWSR1}
\end{equation}

We will often characterize our results in terms of 
the centroid, $\bar S$, and the width, $\Delta S$,
of the transition strengths, defined in terms of $S_1$ and $S_2$
\begin{equation}
\bar S=\frac{S_1}{S_0}, \;\;\,\,\,\,
\Delta S=\sqrt{\frac{S_2}{S_0}-\bar S^2}.
\end{equation}
Both the centroid and the width characterize global properties of collective
excitations.

\subsection{Shell Model}
\label{ssec:SM}

In the interacting SM, the number of possible many-body
configurations is restricted by two means: first, one assumes that only a
limited number of nucleons interact (valence particles), the rest forming
an inert core, and second, the active particles are restricted to a small
number of single particle states (valence space). Usually, the valence
space is restricted to a major oscillator shell. Diagonalization of an
effective Hamiltonian provides the low lying states by means of the
Lanczos algorithm \cite{Whitehead77}; 
the corresponding wavefunctions are eigenstates of the
parity, total angular momentum and isospin.

The reduction of the available single particle states and active particles
makes the diagonalization numerically tractable. There is a downside
though: in order to take into account the restriction of the Hilbert space
one has to use modified (effective) operators to describe observables or
transitions. In many cases a simple phenomenological use of either 
enhanced or quenched couplings, most famously the former for E2 transitions 
and the latter for Gamow-Teller transitions, yields good agreement 
with experiment.

In order to calculate transition strengths in SM, we have used
the Lanczos moment method: the
transition operator is applied on the initial state wavefunction. Then,
the newly obtained state is used as the starting state, or pivot, for 
diagonalization by
means of the Lanczos algorithm. The size of the pivot vector is the 
total transition strength $S_0$, and the overlap of the pivot with 
the final eigenstate, which it turns out can be read off trivially,
is the transition amplitude. The interested reader is referred to 
Ref.~\cite{LanczosMoment} for details.

\subsection{RPA}

Several textbooks \cite{ring,heyde} cover the RPA, so we skip
detailed derivation and just review principal steps.

The starting point of the RPA is a self-consistent mean-field solution, a
Slater determinant which can break symmetries and which ignores correlations.
This determines a deformed particle-hole basis, where the occupation
numbers are zero for particle states and one for hole states.

The RPA ground state is defined as the vacuum for a set of quasiboson operators,
\[
\beta_{\nu}|RPA\rangle=0,
\]
while the excited states, approximate solutions of Eq. (\ref{eq:eig}),
are given by
\[
|\nu\rangle = \beta_\nu^\dagger |RPA\rangle.
\]
And because we assume the excited states to be 1p-1h correlations, one
writes the quasiboson creation operators as \cite{ring}
\begin{equation}
\beta_\nu^\dagger = \sum_{mi}\left(X_{mi}^\nu b_{mi}^\dagger
-Y_{mi}^\nu b_{mi}\right),
\label{exstRPA}
\end{equation}
where $b_{mi}^\dagger\approx c_m^\dagger c_i$ is the approximate boson
mapping of the (deformed) fermion operators.  We use the conventional
notation that $m$, $n$ are unoccupied (particle) states, while $i$,
$j$ are occupied (hole) states, respectively. Note that while in Eq.
(\ref{exstRPA}) the first terms describe particle-hole correlations on top
of the HF state, the terms $Y^\nu_{mi} b_{mi}$ describe correlations
coming from 2p-2h configurations in the ground state. 
In the RPA one assumes that the ground state
is still very close to the HF solution, so that the hole-particle
amplitudes $Y$ are much smaller than the particle-hole amplitudes $X$.
Finally, the eigenvalue equation
\begin{equation}
\left(
\begin{array}{cc}
A & B \\
-B^* & -A^*
\end{array}\right)
\left(
\begin{array}{c}
X\\
Y
\end{array}\right)=
\Omega
\left(
\begin{array}{c}
X\\
Y
\end{array}\right),
\label{RPAmatrix}
\end{equation}
determines the excitation energies $\Omega_\nu$ and the particle-hole
$X$ and hole-particle amplitudes $Y$.
$A$ and $B$ are matrices given by
\begin{equation}
A_{nj,mi} \equiv \left \langle HF  \left |
\left [ \hat{c}^\dagger_j \hat{c}_n , [\hat{H} , \hat{c}^\dagger_m
\hat{c}_i ] \right ]\right | HF  \right \rangle,
\label{defA}
\end{equation}
\begin{equation}
B_{nj,mi} \equiv \left \langle HF  \left |
\left [   [\hat{H} , \hat{c}^\dagger_n
\hat{c}_j ], \hat{c}^\dagger_m \hat{c}_i \right ]
\right | HF  \right \rangle.
\label{defB}
\end{equation}

Thouless showed \cite{thouless61} that if the HF solution corresponds to a
minimum in energy surface, the corresponding RPA equation
(\ref{RPAmatrix}) has only real frequencies. In addition, if the
Hartree-Fock state is invariant under some particle-hole transformation,
such as rotation about an axis, this corresponds to a zero frequency RPA
mode. Thus, the generators of symmetries broken by a mean-field solution are
eigenvectors of Eq. (\ref{RPAmatrix}) lying at at zero excitation energy.
This is frequently interpreted as ``approximate restoration of broken
symmetries'' \cite{ring}; in fact, it is more accurate to say 
that the RPA \textit{respects} symmetries by
separating out exactly spurious motion. In
Sec. \ref{Sec:compstr} we present evidence that significant 
part of the quadrupole response for even-even nuclei is still 
retained in the ground state, which suggests
that broken symmetries are only partially restored by the RPA.

The RPA provides a model for excited states, and to calculate the transition
probability from any non-spurious state to the ground state one needs the
transition matrix element $\langle \nu |F|{\rm RPA} \rangle$. In the RPA, the
latter can be written in terms of particle-hole amplitudes $X^\nu$ and
$Y^\nu$, namely, if one has a 
one-body transition operator (and a spherical tensor 
of rank $K$),  $F_{KM}$, which can be written as 
\begin{equation}
F_{KM}=\sum_{mi}(f^M_{mi}c_m^\dagger c_i + (-1)^M f_{mi}^{-M} c_i^\dagger c_m),
\label{trans1body}
\end{equation}
then \cite{ring}
\begin{equation}
\langle \nu |F_{KM}| \mathrm{RPA} \rangle =
  f^M \cdot X^\nu +
(-1)^Mf^{-M} \cdot Y^\nu
\label{transMErpa}
\end{equation}
where $f\cdot X = \sum_{mi} f_{mi} X_{mi}$ etc.

With the transition matrix element  (\ref{transMErpa}), it is possible to
calculate in the RPA any moment of the distribution strength, and
therefore the total strength, the centroid and the width. 
Sec.~\ref{Sec:compstr} compares the SM and RPA predictions for these
quantities, as well as individual transition strengths in several nuclei.
Before proceeding with our numerical results, there are two points to discuss. 

First, the energy-weighted sum rule $S_1$.
The RPA has the famous property 
\cite{ring,thouless61}:
\begin{equation}
\sum_{\nu}\Omega_\nu |\langle \nu |F|{\rm RPA} \rangle|^2 =
\frac{1}{2}\langle {\rm HF} | [F ,[H,F]]| {\rm HF} \rangle.
\label{ewsr1HF}
\end{equation}
In the Appendix we revisit the derivation of 
Eq.~(\ref{ewsr1HF}) and find that it can be violated 
\textit{if} an exact symmetry such as rotational invariance 
is broken. In Sec.~\ref{sec:ewsr} we confirm the violation numerically, 
and find the worse case to be where the bulk of the 
transition strength lies very low in energy, such as isoscalar E2.
This appears to be a new result. Because applications of RPA have usually 
assumed spherical symmetry they are not invalidated, but ambitious 
RPA calculations that allowed for broken symmetries in the 
mean-field \cite{bertsch2000} should be approached with caution. 
Of particular interest, which we have not yet explored, is the QRPA 
which breaks particle number conservation.

A second point we would like to discuss involves ground-state to
ground-state transitions. (These  
are nothing more than ground-state expectation 
values of non-scalar operators. We discuss ground state expectation 
values of \textit{scalar} operators in \cite{johnson2002}; in 
principle one can extend such calculations to non-scalar operators, but 
we have not yet done so. Marshalek and Weneser \cite{weneser} 
discuss expectation values for electric quadrupole and magnetic moments, 
but their approach is not very transparent for general implementation.) 
 Most discussions of the RPA do not give a
well-defined procedure to calculate
such transitions, in part because they vanish when the Hartree-Fock 
state has spherical symmetry.
While such g.s.~to g.s.~transitions are forbidden for a spherical (hence $J=0$) 
state, they are in general not forbidden for nonspherical HF states.  
If RPA does not fully restore
broken symmetries, a significant contribution to the total strength could 
be absorbed into otherwise forbidden g.s.~to g.s.~transitions.
We investigate in detail this
point in Sec.~\ref{sec:downEn}, and indeed we find that 
significant strength to excited states can be missing for even-even nuclides.

\section{RPA vs. exact shell-model strengths}
\label{Sec:compstr}

In order to test the RPA reliability for computing transition strengths,
we calculate both the mean-field and exact solutions in the same model
space, using the same Hamiltonian.

\subsection{Model space, interaction and transition operators}

We work in full $0\hbar\omega$ shell model spaces, restricting the single
particle states to one major shell. Most of our examples were computed 
in the $sd$ shell, limiting the nucleons outside an inert $^{16}$O core to
the single particle orbits $1s_{1/2}$-$0d_{3/2}$-$0d_{5/2}$. Additionally, we
considered two nuclei, $^{44}$Ti and $^{46}$V, in the $pf$ shell, i.e.,
$1p_{1/2}-1p_{3/2}-0f_{5/2}-0f_{7/2}$ single particle states outside
$^{40}$Ca core. For the interaction, we used the Wildenthal ``USD'' in the
$sd$ shell \cite{wildenthal} and the monopole-modified Kuo-Brown ``KB3''
in the $pf$ shell \cite{KB3}. We emphasize that due to our restriction to
a single major shell and limitation to mixing angular degrees of freedom,
the mean-field solution can break only the rotational symmetry.

For testing purposes, we have considered $F_{JT}=\tilde e_Tr^JY_J$, with
$J=2$, that is isoscalar ($T=0$) and isovector ($T=1$) electric quadrupole (E2).
While  good agreement with experimental transitions strengths requires 
nontrivial effective
proton and neutron charges, the main contribution of effective charges 
is a rescaling of the
strengths; therefore for simplicity we took the bare charges,
$e_p =1$ and $e_n = 0$. This might appear to suggest that we only 
considered the proton response, but we computed the isoscalar and 
isovector responses separately, that is, incoherently (only if 
the isoscalar and isovector responses were summed coherently would one 
have pure proton response). 

In addition, we tested transition distributions for spin flip (SF) and
Gamow-Teller (GT) which are the isoscalar and isovector components of the
spin operator $\sigma$.  
To avoid confusion, note that the actual GT operator we
used is $\mathbf{\sigma} \tau_z$ so that $T_z=Z-N$ was conserved.

A large fraction application of RPA calculations are to E1 transitions. Because
our shell model valence space does not include single-particle states
of opposite parity, we could not investigate E1 transitions here.

\subsection{Results for isovector quadrupole, SF, and GT
transition operators}
\label{sec:upEn}

In this section we show results for isovector E2, SF (spin-flip) and GT 
(Gamow-Teller) transition
operators. The main common feature is that their collective transitions
lie relatively high in energy. We find that for such transitions the RPA
is in reasonably good agreement with the SM results, especially for the
total transition strength.

\begin{figure}
        \centering
	\includegraphics*[scale=0.60]{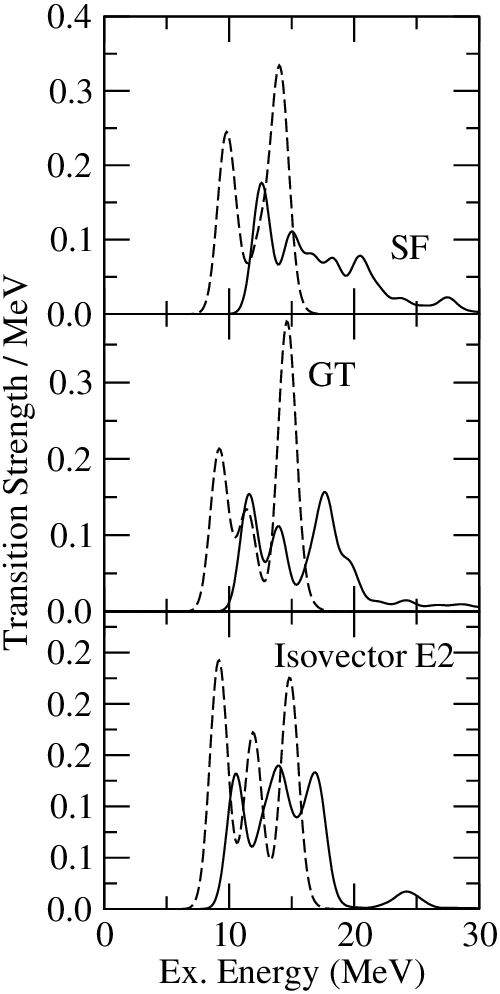}
	\caption{Isovector E2, SF and GT transition strengths for $^{20}$Ne.
        Both the exact SM (solid curve) and RPA (dashed curve) distributions
        have been smoothed with a  Gaussian of width 0.7 MeV to
        facilitate comparison.}
	\label{fig:Ne20IV}
\end{figure}

\begin{figure}
        \centering
	\includegraphics*[scale=0.60]{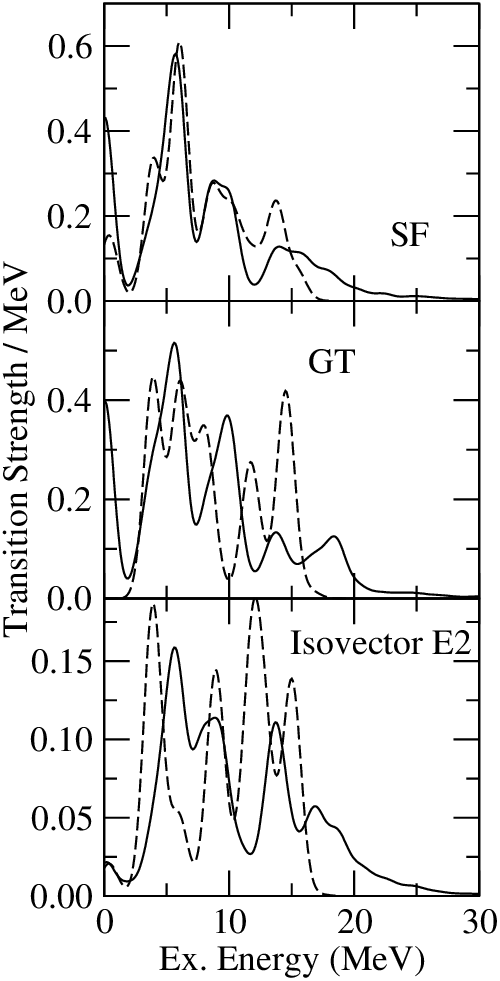}
	\caption{Same as in Fig. \ref{fig:Ne20IV} for $^{21}$Ne.}
	\label{fig:Ne21IV}
\end{figure}

\begin{figure}
        \centering
	\includegraphics*[scale=0.60]{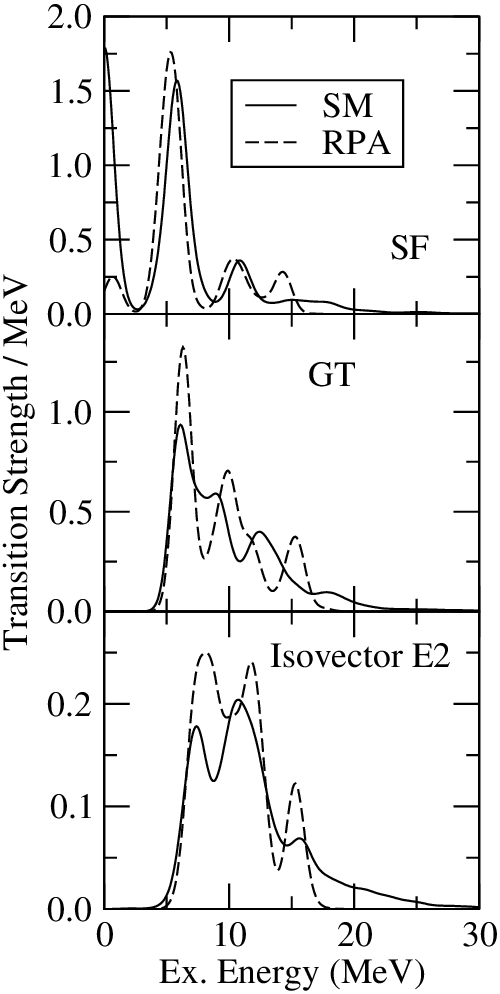}
	\caption{Same as in Fig. \ref{fig:Ne20IV} for $^{22}$Na.}
	\label{fig:Na22IV}
\end{figure}

Figures \ref{fig:Ne20IV}-\ref{fig:Na22IV} compare the RPA and SM
transition strengths; we choose for exemplification $^{20}$Ne (even-even),
$^{21}$Ne (even-odd) and $^{22}$Na (odd-odd), but the general trend is the
same for all the nuclides investigated. The excitation spectra are discrete, 
but to guide the eye we folded in a Gaussian of width 0.7 MeV. 
In addition, tables
\ref{table:E2IV}-\ref{table:GT} summarize the results in both SM and RPA
for several nuclei; we present only the total strengths, the centroids and
the widths of the distributions.

The figures show that the RPA calculations follow the general features of
the SM transition strength distributions. Note however that by comparison
to SM, the RPA distributions have smaller
widths (see tables \ref{table:E2IV}-\ref{table:GT}). This is not
surprising, as higher-order particle-hole correlations are expected to
further fragment the distribution. The RPA centroids are generally shifted to
lower energies than the SM.  Although the centroids are related 
to the energy-weighted sum rule $S_1$, we remind the reader that 
we do not violate Eq.~(\ref{genEWSR1}) because the HF state is only an 
approximation to the ground state. Furthermore, the shift in the centroid does
not appear correlated with the correctness
of the RPA estimations of the ground state energy \cite{stetcu2002} or
other observables \cite{johnson2002}. One might expect that the correct
inclusion of the pairing interaction by means of HFB+QRPA would improve the
results. This is reasonable and worth trying, but 
see discussion and caveats regarding pairing and QRPA in 
\cite{stetcu2002,johnson2002}.

For computational simplicity, we restrict ourselves to real wavefunctions;
this has no effect for even-even nuclei. But because the rotations about
$x$ or $z$ axis are complex, for odd-odd or odd-$A$ nuclei the RPA does
not identify all the corresponding generators as exactly zero-frequency modes. Instead,
we obtain a `soft' mode at very low excitation energy. Transition strengths 
to the soft mode are in fact ground-state-to-ground-state strength normally 
not computed in RPA.

\begin{table}
\caption{Total strength $S_0$, centroid $\bar S$, and width $\Delta S$ for
isovector E2 transition operator. The nuclei have been grouped into
even-even, odd-odd and odd-A.}
\label{table:E2IV}
\begin{ruledtabular}
\begin{tabular}{ccccccc}
 & \multicolumn{2}{c}{$S_0$} & \multicolumn{2}{c}{$\bar S$ (MeV)}
 & \multicolumn{2}{c}{$\Delta S$ (MeV)}\\
\cline{2-3}\cline{4-5}\cline{6-7}
Nucleus &  SM & RPA & SM & RPA & SM & RPA  \\
\hline
$^{20}$Ne &  0.98 & 1.15 & 14.53 & 11.92 & 3.47 & 2.44 \\
$^{22}$Ne &  2.37 & 1.86 &  8.15 &  7.70 & 5.85 & 4.25\\
$^{24}$Mg &  1.88 & 1.96 & 14.40 & 11.86 & 4.09 & 2.46\\
$^{28}$Si &  2.28 & 1.96 & 14.35 & 13.41 & 4.29 & 1.93\\
$^{36}$Ar &  1.38 & 1.34 & 12.49 & 11.01 & 4.24 & 3.56\\
$^{44}$Ti &  2.15 & 1.90 &  8.23 &  6.68 & 2.86 & 1.98\\
\hline
$^{22}$Na &  1.60 & 1.69 & 11.67 & 10.44 & 4.37 & 2.61\\
$^{24}$Na &  2.07 & 2.09 &  9.82 &  8.21 & 6.26 & 4.14\\
$^{46}$V  &  2.32 & 3.00 &  7.96 &  6.62 & 4.17 & 1.97\\
\hline
$^{21}$Ne &  1.39 & 1.43 & 10.52 &  9.41 & 5.66 & 4.27\\
$^{25}$Mg &  2.28 & 2.20 & 11.48 &  9.47 & 6.21 & 4.41\\
$^{29}$Si &  2.52 & 2.20 & 11.61 & 10.21 & 5.68 & 4.25
\end{tabular}
\end{ruledtabular}
\end{table}

\begin{table}
\caption{Same as in Table \ref{table:E2IV} for SF transition operator.}
\label{table:SF}
\begin{ruledtabular}
\begin{tabular}{ccccccc}
 & \multicolumn{2}{c}{$S_0$} & \multicolumn{2}{c}{$\bar S$ (MeV)} &
\multicolumn{2}{c}{$\Delta S$ (MeV)} \\
\cline{2-3}\cline{4-5}\cline{6-7}
Nucleus &  SM & RPA & SM & RPA & SM & RPA  \\
\hline
$^{20}$Ne &  1.05 & 1.23 & 17.10 & 12.26 & 4.38 & 1.95\\
$^{22}$Ne &  3.53 & 4.44 & 11.40 &  8.82 & 4.32 & 2.37\\
$^{24}$Mg &  4.15 & 4.78 & 13.22 & 10.17 & 4.48 & 1.97\\
$^{28}$Si &  5.82 & 5.20 & 12.75 & 11.62 & 4.34 & 1.89\\
$^{36}$Ar &  2.68 & 2.70 & 14.53 & 11.17 & 3.69 & 2.88\\
$^{44}$Ti &  2.56 & 3.32 &  9.98 &  7.86 & 2.56 & 1.55\\
\hline
$^{22}$Na &  8.57 & 5.78 &  5.01 &  6.67 & 5.22 & 3.49\\
$^{24}$Na & 10.06 & 7.66 &  5.83 &  7.05 & 5.48 & 3.19\\
$^{46}$V  &  5.44 & 7.68 &  8.76 &  6.40 & 2.51 & 2.34\\
\hline
$^{21}$Ne &  4.02 & 3.54 &  7.50 &  7.62 & 5.92 & 3.95\\
$^{25}$Mg &  6.94 & 6.33 &  9.19 &  8.47 & 5.73 & 3.36\\
$^{29}$Si &  8.42 & 8.47 &  9.38 &  7.86 & 5.07 & 4.45
\end{tabular}
\end{ruledtabular}
\end{table}

\begin{table}
\caption{Same as in Table \ref{table:E2IV} for GT transition operator.}
\label{table:GT}
\begin{ruledtabular}
\begin{tabular}{ccccccc}
 & \multicolumn{2}{c}{$S_0$} & \multicolumn{2}{c}{$\bar S$ (MeV)}
 & \multicolumn{2}{c}{$\Delta S$ (MeV)} \\
\cline{2-3}\cline{4-5}\cline{6-7}
Nucleus &  SM & RPA & SM & RPA & SM & RPA  \\
\hline
$^{20}$Ne & 1.05 & 1.33 & 16.32 & 12.53 & 4.35 & 2.42 \\
$^{22}$Ne & 3.87 & 4.85 & 12.00 &  9.37 & 4.48 & 3.16 \\
$^{24}$Mg & 4.26 & 4.85 & 14.46 & 11.74 & 4.24 & 2.42 \\
$^{28}$Si & 6.65 & 5.70 & 15.19 & 13.77 & 3.59 & 1.88 \\
$^{36}$Ar & 2.74 & 2.79 & 14.85 & 12.09 & 3.45 & 2.99 \\
$^{44}$Ti & 3.03 & 3.74 & 10.12 &  8.42 & 2.86 & 2.43 \\
\hline
$^{22}$Na & 5.51 & 5.47 &  9.96 &  9.28 & 4.35 & 3.18 \\
$^{24}$Na & 7.43 & 7.71 & 10.32 &  9.29 & 4.87 & 3.48 \\
$^{46}$V  &10.60 & 7.85 &  4.93 &  8.15 & 4.37 & 2.28 \\
\hline
$^{21}$Ne & 4.25 & 3.55 &  7.87 &  8.67 & 5.97 & 3.98 \\
$^{25}$Mg & 7.12 & 6.76 & 11.02 & 10.00 & 6.05 & 4.21 \\
$^{29}$Si & 9.42 & 8.63 & 12.28 & 10.39 & 5.41 & 4.99
\end{tabular}
\end{ruledtabular}
\end{table}

To summarize the results in this section, we have compared the SM and RPA
distribution strengths for isovector E2, SF and GT transition operators.
We found in general good agreement for the total strength in several
nuclei. While less satisfactory, the centroids and widths of the
distributions are still close. As a general feature however, the RPA
distributions are smaller in width and lower in energy than the SM results.


\subsection{Results for isoscalar quadrupole response}
\label{sec:downEn}

This section presents comparison between the SM and RPA distribution
strength for the isoscalar quadrupole transition operator. The main
difference with respect to the other transitions investigated in this
paper is that the collective strength lies very low in energy, for
realistic Hamiltonians have a strong attractive isoscalar
quadrupole-quadrupole component.

\begin{figure}
  \centering
	\includegraphics*[scale=0.60]{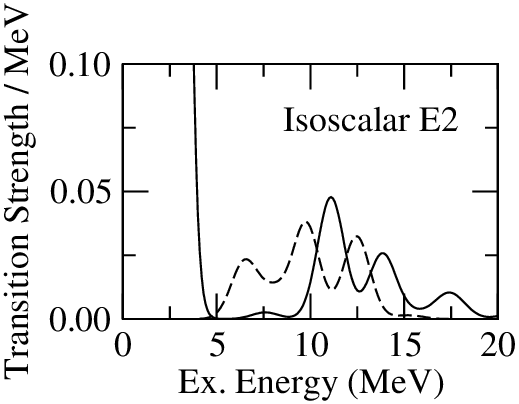}
	\caption{Isoscalar E2 transition strengths for $^{20}$Ne.
        The SM (solid curve) and RPA (dashed curve) distributions
        have been smoothed with a Gaussian of width 0.7 MeV.  The 
        large collective peak at low but nonzero excitation energy
	for the SM is absent in the RPA; see text for discussion.}
	\label{fig:Ne20IS}
\end{figure}

\begin{figure}
  \centering
	\includegraphics*[scale=0.60]{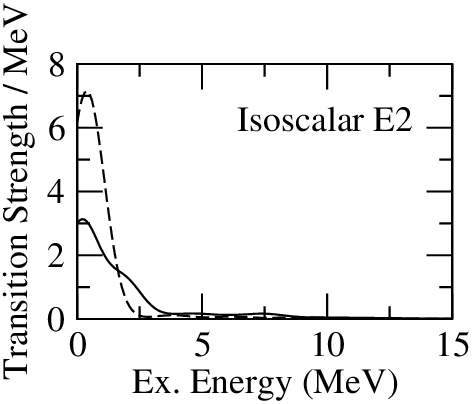}
	\caption{Same as Fig. \ref{fig:Ne20IS} for $^{21}$Ne.}
	\label{fig:Ne21IS}
\end{figure}

\begin{figure}
  \centering
	\includegraphics*[scale=0.60]{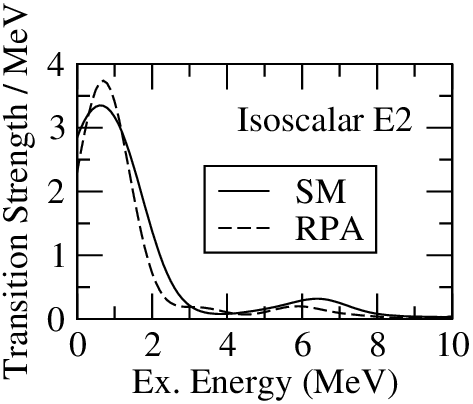}
	\caption{Same as Fig. \ref{fig:Ne20IS} for $^{22}$Na.}
	\label{fig:Na22IS}
\end{figure}

We considered again for comparison the same nuclides investigated
previously, and we plot the SM and RPA distributions in figures
\ref{fig:Ne20IS}-\ref{fig:Na22IS}. Characteristics of the distributions
for several other nuclei are given in table \ref{table:E2IS}. In contrast
with the results in Sec.~\ref{sec:upEn}, we find a large discrepancy
between the total strengths in RPA and SM, especially for even-even
nuclei.

Figure \ref{fig:Ne20IS} shows that, if one ignores the low energy
transitions, one obtains again a reasonable agreement between the SM and RPA
distributions. Similar features encountered for other transitions appear,
that is a lower energy centroid and smaller width of the RPA
distribution with respect to SM.

\begin{table}
\caption{Same as in Table \ref{table:E2IV} for
isoscalar E2 transition operator.}
\label{table:E2IS}
\begin{ruledtabular}
\begin{tabular}{ccccccc}
 & \multicolumn{2}{c}{$S_0$} & \multicolumn{2}{c}{$\bar S$ (MeV)}
 & \multicolumn{2}{c}{$\Delta S$ (MeV)} \\
\cline{2-3}\cline{4-5}\cline{6-7}
Nucleus &  SM & RPA & SM & RPA & SM & RPA  \\
\hline
$^{20}$Ne &  7.86 & 0.19 &  2.12 &  9.81 & 1.92 &  2.30 \\
$^{22}$Ne &  9.36 & 0.89 &  2.01 &  5.52 & 2.19 &  2.79 \\
$^{24}$Mg &  12.57& 0.51 &  2.13 &  7.99 & 2.09 &  2.75\\
$^{28}$Si &  12.04& 0.56 &  2.51 &  9.88 & 2.33 &  2.33 \\
$^{36}$Ar &  7.17 & 0.23 &  2.42 &  9.57 & 1.91 &  2.74\\
$^{44}$Ti &  10.87& 1.50 &  1.73 &  3.99 & 1.73 &  1.70\\
\hline
$^{22}$Na &  9.53 & 7.49 &  1.47 &  1.27 & 2.63 &  1.82 \\
$^{24}$Na &  8.81 & 6.33 &  2.10 &  1.81 & 2.85 &  1.88\\
$^{46}$V  &  15.21&15.20 &  1.62 &  0.87 & 1.94 &  1.63\\
\hline
$^{21}$Ne &  8.74 &13.27 &  1.53 &  0.64 & 2.82 &  1.35 \\
$^{25}$Mg & 10.71 &12.49 &  2.25 &  1.08 & 2.66 &  1.62 \\
$^{29}$Si &  9.70 & 1.38 &  2.72 &  4.66 & 2.62 & 4.25
\end{tabular}
\end{ruledtabular}
\end{table}

As for the relative good agreement for odd-odd and odd-$A$ nuclei, we have
to point out that most of the RPA strength is concentrated in the lowest
energy state which, as already noted, appears just as an artifact of our
approach (restriction to real numbers). A full treatment of
rotations by inclusion of complex numbers would shift these `soft-mode' states to
zero modes, that is, degenerate with respect to the ground-state, and 
we would expect the odd-odd and odd-$A$ cases to then resemble the even-even 
cases: missing the low-energy collective strength. (Note that qualitatively the
results for $^{29}$Si, for which we obtain the correct number of zero RPA
modes, are similar to the even-even nuclei.)
Conversely, we can turn around these results into a hypothesis: that the 
missing low-lying collective strength in even-even nuclides are due to 
incomplete symmetry restoration, and that the missing strength resides in 
the RPA ground state.  The fact that the missing strength shows up in 
soft modes that arise as artifacts of our computational methods bolsters 
this hypothesis. For the interested reader, more details can be found 
in the next section. 

\subsection{``Missing strength'' and broken symmetries}
\label{sec:ewsr}

In this section we provide further evidence supporting our hypothesis 
that the low-lying collective strength is missing 
due to incomplete symmetry restoration in the RPA, and a 
significant fraction of the RPA strength
gets absorbed in a ground state to ground state transition.

Our first test of incomplete symmetry restoration is the comparison of 
the transition strength for spherical and deformed HF solutions. 
While the proton-neutron interaction induces deformation in
the HF Slater determinant for $^{28}$Si, it is possible nevertheless to
force a transition to a spherical HF state (both
protons and neutrons filling the $d_{5/2}$ orbits) by increasing the gap
between the $d_{5/2}$ single particle energy and the other
single-particle states. The above-reported values for $^{28}$Si 
use the USD value of $\epsilon(d_{5/2}) = -3.94$ MeV, which yields a 
deformed HF state.  In addition 
we computed $^{28}$Si at $\epsilon(d_{5/2}) = -5.64$ MeV and $-5.74$ MeV. 
At $-5.64$ MeV the HF state is still deformed while at $-5.74$ MeV the HF state is 
spherical. (Actually, for these values of $\epsilon(d_{5/2})$ there exist  
both spherical and deformed locally stable HF solutions, but at $-5.64$ MeV 
the deformed state has a slightly lower HF energy while at $-5.74$ MeV the spherical 
state has the lowest HF energy. Thus this is a first order `phase transition' 
as described in section 4 of \cite{thouless61}. The so-called collapse or 
breakdown of RPA, readily seen in toy models 
such as the Lipkin model \cite{ring}, only occurs when one has a 
second order `phase transition,' when one has only one stable HF solution. 
In other words, our RPA calculations do not collapse at the transition point.)

\begin{figure}
  \centering
	\includegraphics*[scale=0.60]{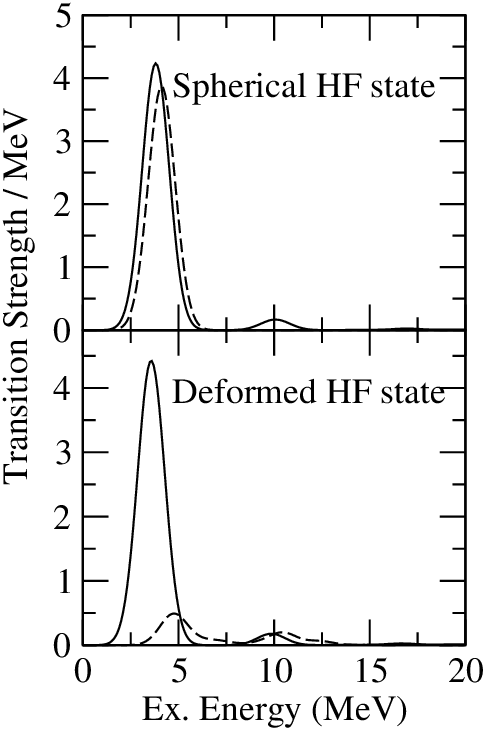}
	\caption{$^{28}$Si: Isoscalar E2 for deformed and spherical HF state.}
	\label{fig:Si28phT}
\end{figure}

Figure \ref{fig:Si28phT} shows small difference in
the SM strength distribution in contrast with a dramatic change for RPA.
The difference between the $d_{5/2}$ single-particle energies in the two
cases is small and one can follow a smooth change for all observables
in the SM; we have therefore no reason to suspect any fundamental
difference in the structure of the states. Note however that, when the HF
state is spherical, the low lying states are correctly described in the
RPA, the reason why the RPA was successful in describing low lying
collectivity in closed shell nuclei; but note also that 
the high-lying part of the strength is not correctly described.
 In contrast, when the HF state is
deformed, the RPA strength distribution changes dramatically, even 
through the SM strength distribution does not: the low-lying strength 
vanishes, but the high-lying strength is approximately correct.

As a second test, we compare  the total strength $S_0$ and the energy-weighted 
sum rule $S_1$ computed in different ways. 

Table \ref{table:s0} presents the total strength $S_0$ for a transition 
operator $F$, where $F$ is either the spin-flip operator or isoscalar E2 
operator. The columns labled `SM' are the exact shell model results, 
for which $S_0 = \langle 0 | F^2 | 0 \rangle =\sum_\nu |\langle \nu | F | 0 \rangle |^2 $.
Of course, for the shell model both methods 
yield the same result. The columns RPA-X and RPA-$\Sigma$ correspond to 
equivalent methods for the RPA. RPA-X is the expectation value 
$\langle RPA | F^\dagger F | RPA \rangle$ as laid out in \cite{johnson2002}, 
where we showed that the RPA expectation value was often a reasonable 
approximation to the shell model result, though not always as 
seen for some of the spin-flip cases. RPA-$\Sigma$ is the 
sum $\sum_\nu |\langle \nu | F | RPA \rangle |^2 $, where the sum 
is only to excited states as g.s.-to-g.s. transitions are difficult to 
define in RPA.  

The horizontal lines in Table \ref{table:s0} segregate 
even-even nuclei with deformed HF states, even-even spherical nuclei, 
odd-odd nuclei, and odd-A nuclei. Keep in mind that for the latter two groups 
we do not get all the true zero modes (because we restricted the Slater determinant to 
real single-particle wavefunctions), but at least one zero mode is 
replaced by a soft mode, except for the case of 
$^{29}$Si which does have all the expected zero modes. 

What do we learn from Table \ref{table:s0}? We draw the reader's attention
to the isoscalar E2 strength in deformed even-even-nuclides, and in 
$^{29}$Si, all of which have the expected number of exact zero modes.
Here the summed RPA
strength (RPA-$\Sigma$) is dramatically and consistently smaller than 
either the exact SM result, or the expectation value RPA-X. By way of 
contrast, the nuclides with spherical HF states and thus no zero modes, 
or those who have soft modes rather than zero modes, 
have summed RPA strength in reasonable accord with 
the SM total strength. Furthermore the RPA expectation value of 
$Q\cdot Q$ also agrees with the SM total strength, which suggests to us 
that some of the  missing RPA strength is in g.s. to g.s. transitions. 
This line of reasoning is weakened by the poor reliability of the RPA 
expectation value, as discussed in \cite{johnson2002} and as seen in the 
spin-flip values, which take on unphysical negative values for spherical nuclei.

\begin{table}

\caption{Comparison total strength $S_0$ as computed 
in the shell model (SM), by taking the expectation value 
of an observable (RPA-X), and by summing the RPA strengths
directly (RPA-$\Sigma$); see text for more details. 
The horizontal lines separate even-even deformed nuclei, 
even-even spherical nuclei, odd-odd and odd-A. 
Notation on $^{28}$Si: $^\dagger$ indicates 
$\varepsilon_{d5/2}=-5.74$ MeV and a 
spherical HF state.}
\label{table:s0}
\begin{ruledtabular}
\begin{tabular}{ccccccc}
 & \multicolumn{3}{c}{SF} & \multicolumn{3}{c}{Isoscalar E2}\\
\cline{2-4}\cline{5-7}
Nucleus &  SM & RPA-X & RPA-$\Sigma$ &
SM & RPA-X & RPA-$\Sigma$  \\
\hline
$^{20}$Ne & 1.05 & 1.34 & 1.23 & 7.86 & 8.17 & 0.19 \\
$^{22}$Ne & 3.52 & 1.94 & 4.44 & 9.36 & 9.98 & 0.89 \\
$^{24}$Mg & 4.15 & 5.53 & 4.78 &12.57 & 12.52& 0.51 \\
$^{36}$Ar & 2.68 & 2.90 & 2.70 & 7.17 &  7.66& 0.23 \\
$^{28}$Si & 5.82 & 5.13 & 5.20 &12.04 & 13.77& 0.56 \\
\hline
$^{28}$Si$^\dagger$ & 9.07 & -5.22 & 12.53& 7.95& 7.50 & 6.86 \\
$^{22}$O & 5.04 & -0.60 & 6.31 & 2.71 & 2.75 & 2.36 \\
$^{24}$O & 5.17 & -1.07 & 6.24 & 1.88 & 1.92 & 1.57 \\
\hline
$^{22}$Na & 8.57 & 7.64 & 5.78 & 9.53 & 9.92 & 7.49 \\
$^{24}$Na & 10.06& 6.97 & 7.66 & 8.81 & 9.61 & 6.33 \\
\hline
$^{21}$Ne & 4.02 & 1.77 & 3.54 & 8.74 & 8.69 & 13.27\\
$^{25}$Mg & 6.94 & 1.52 & 6.33 &10.71 & 12.34& 12.49 \\
$^{29}$Si & 8.42 & 5.77 & 8.47 & 9.70 & 11.67&  1.38
\end{tabular}
\end{ruledtabular}
\end{table}

Therefore, to further dissect this issue, in Table \ref{table:s1} we consider 
the energy-weighted sum rule $S_1$. The SM value is 
$\sum_\nu (E_\nu -E_0) |\langle \nu| F | 0 \rangle |^2$. The RPA value is 
$\sum_\nu \Omega_\nu |\langle \nu | F | { \rm RPA} \rangle |^2$, while the HF value 
is $\frac{1}{2} \langle {\rm HF} | [ F,[H,F] | {\rm HF} \rangle$ 
(for technical reasons, discussed in the Appendix, 
we can only compute this for even-even nuclei).  

For nuclei with spherical HF states, that is, no zero modes, the RPA $S_1$ 
and the HF value are identical; this is the usual theorem regarding the energy 
weighted-sum rule.  For nuclei with deformed HF states, and thus with 
zero modes, the RPA and HF values differ, a small amount for isovector E2 and 
dramatically for isoscalar E2. (We also computed but do not show values for the 
spin-flip operator; the results are similar to isovector E2, although 
with an even smaller contribution 
from the zero modes.)  Overall these results are consistent with our 
hypothesis that low-lying strength is subsumed into the RPA ground state 
due to incompletely restoration of a broken symmetry; the difference is larger 
for isoscalar E2 because of the large strength at low energy.

\begin{table}

\caption{Comparison of energy-weighted sum rule $S_1$ as computed 
in the shell model (SM), taking the weighted sum of 
RPA strengths (RPA), and taking the HF expectation value
of a double-commutator (HF); see text for more details. 
Notation on $^{28}$Si, same as 
for Table \ref{table:s0}.}
\label{table:s1}
\begin{ruledtabular}
\begin{tabular}{ccccccc}
  & \multicolumn{3}{c}{Isovector E2}&
 \multicolumn{3}{c}{Isoscalar E2}\\
\cline{2-4}\cline{5-7}
Nucleus &  SM & RPA & HF & SM & RPA & HF \\
\hline
$^{20}$Ne & 14.27 & 13.74 & 13.74 &  16.63 & 1.82 & 7.43 \\
$^{22}$Ne & 19.28 & 14.34 & 14.40 &  18.84 & 4.92 & 10.51 \\
$^{24}$Ne & 21.89 & 14.89 & 15.06 &  20.87 & 8.42 & 11.99 \\
$^{24}$Mg & 27.09 & 23.28 & 23.28 &  26.71 & 4.08 & 14.87 \\
$^{36}$Ar & 17.24 & 14.72 & 14.72 &  17.32 & 2.21 & 8.64 \\
$^{28}$Si & 32.66 & 26.22 & 26.22 &  30.22 & 5.58 & 17.67 \\
\hline
$^{28}$Si$^\dagger$ & 40.13 & 35.76 & 35.76 & 34.31 & 28.26 & 28.26 \\
$^{22}$O & 11.46 & 8.56 & 8.56 &  11.46 & 8.56 & 8.56 \\
$^{22}$O & 10.42 & 7.99 & 7.99 &  10.42 & 7.99 & 7.99 \\
\end{tabular}
\end{ruledtabular}
\end{table}

\section{conclusions}
\label{Sec:concl}

The purpose of this paper was to investigate the reliability of the RPA
for calculating transition strengths in nuclei. To accomplish this we have
computed the RPA and shell model strength distributions in the same $0\hbar\omega$
shell model space.

The comparison between RPA and SM showed two different results, depending
upon the nature of transitions. Thus, we found that when the
strong collectivity lies at high energies, such as isovector E2, SF and GT
transitions, the RPA and SM are in reasonable agreement. When the
transitions lie at low energies however, the agreement is poor. We
presented evidence that the problem arises from an incomplete
restoration of the symmetries broken by the mean-field; for low-lying
transitions we propose that significant part of the
transition strength is subsumed into the RPA ground state. Future work should
directly investigate  ground-state to ground-state transitions in the RPA.  
(These are also needed for ground state moments, such as magnetic dipole or 
electric quadrupole, of odd-A nuclides.) 
Finally, we also have found, both analytically and numerically, 
that the standard lore regarding the RPA 
energy-weighted sum rule Eq.~(\ref{ewsr1HF}) does not hold if an exact 
symmetry is broken, particularly if the centroid of the transition 
strength is very low in energy.

This paper also marks a final stage within a larger project to test
the reliability of the HF+RPA for a global microscopic theory of nuclear
properties \cite{stetcu2002,johnson2002}. 
We conclude that HF+RPA is a good starting point 
for such a task, but because of occasional failures 
future work should investigate Hartree-Fock-Bogoliubov 
+QRPA and extensions such as renormalized RPA, self-consistent
RPA, etc., (see Ref.~\cite{ring} as well as the bibliographies 
of \cite{stetcu2002,johnson2002}) and the second RPA, which has been 
shown to differ significantly from the standard RPA in its description of 
E2 giant resonances of $^{16}$O \cite{secondRPA}.
 Our work suggests an 
important and specific test of any such ``improvement'' to RPA: the description of 
low-lying collective strength, such as isoscalar E2, 
which is sensitive to restoration of the rotational symmetry broken 
by a deformed mean-field state.

\appendix*

\section{The energy-weighted sum rule, revisited}

In Table \ref{table:s1} we saw a discrepancy 
between two ways to compute the energy-weighted sum rule, 
\begin{equation}
S_1^{HF} = \frac{1}{2}\sum_{M}(-1)^M\langle \mathrm{HF} | [F_{K,-M} ,
[H,F_{KM}]]|\mathrm{HF} \rangle
\end{equation}
and 
\begin{equation}
S_1^{RPA} = \sum_\nu \sum_M
\Omega_\nu \left | \langle \nu | F_{KM} | \mathrm{RPA} \rangle \right |^2.
\end{equation}
In textbooks \cite{ring} one finds ``proof'' that $S_1^{HF} = S_1^{RPA}$, 
which is often stated that RPA preserves the energy-weighed sum rule. 
In this Appendix we revisit the proof, with special attention to zero 
modes that arise from broken exact symmetries, and we find that instead  
$S_1^{HF} = S_1^{RPA}+$ a term that arises from zero modes.

Suppose we have a broken symmetry, such as rotational invariance. 
The Hartree-Fock state is deformed and has a particular orientation, 
but the Hartree-Fock energy is independent of the orientation. 
This shows up in the RPA matrix equation (\ref{RPAmatrix}) as a zero-frequency 
mode. For $\Omega > 0$ one has the normalization 
$\vec{X}^2 -\vec{Y}^2 =1$ but this 
normalization is impossible for $\Omega = 0$. 
Instead one introduces  
collective coordinates $\vec{Q}_\nu$ and conjugate momenta 
$\vec{P}_\nu$\cite{ring,weneser}, which satisfy 
\begin{eqnarray}
\textbf{A} \vec{P}_\nu - \textbf{B} \vec{P}^*_\nu = iM_\nu \Omega^2_\nu \vec{Q}_\nu, 
\nonumber \\
\textbf{A} \vec{Q}_\nu - \textbf{B} \vec{Q}^*_\nu = - \frac{i}{M_\nu}
\vec{P}_\nu.
\label{RPAqp}
\end{eqnarray}
Here $M_\nu$ is a constant, interpretable as mass or moment of inertia 
fixed by the normalization of $P, Q$: 
\begin{equation}
\vec{Q}^*_\lambda \cdot \vec{P}_\nu - \vec{Q}_\lambda \cdot \vec{P}^*_\nu
= i \delta_{\lambda \nu}.
\label{qpnorm}
\end{equation}
Note that if \textbf{A} and \textbf{B} are real, then $X$, $Y$ are real, 
but of necessity $P$ and $Q$ are complex (one is real and the other imaginary). 
With these zero-mode frequencies one must supplement the quasiboson 
operators $\beta, \beta^\dagger$ in Eq.~(\ref{exstRPA})
with the generalized coordinate and 
momentum operators ${\cal Q}, {\cal P}$. 

Because of the expansion (\ref{trans1body}) 
one can use the definitions (\ref{defA}),
(\ref{defB}), and
able to use $A$ and $B$ to write $S_1^{HF}$ as
\begin{equation}
\sum_M\sum_{mi,nj}\left(A_{mi,nj}f_{mi}^Mf_{nj}^M-(-)^M B_{mi,nj}f_{mi}^Mf_{nj}^{-M}
 \right) \nonumber
\end{equation}

The matrices $A$ and $B$ can be written in terms of particle hole amplitudes
$X$ and $Y$, and the canonical momentum operators associated with broken
symmetries
\begin{eqnarray}
A=X\Omega X^\dagger + Y^*\Omega Y^T + P M^{-1} P^\dagger,
\\
-B=X\Omega Y^\dagger + Y^*\Omega X^T - P M^{-1} P^T.
\end{eqnarray}
Substitution and some algebra yields
\begin{eqnarray}
\lefteqn{S_1^{HF}
 =S_1^{RPA}
  +} \\
& &  \sum_M\sum_{\mu (\Omega_\mu= 0)}\frac{1}{2M_\mu}
\left | f^M \cdot  P_\mu
 -(-)^M f^{-M}\cdot P_\mu^* \right|^2  \nonumber
\label{eq:corrCom}
\end{eqnarray}
Although we do not show it, one can write the right-hand side
as a double-commutator of boson operators.

As further motivation, one can start from
Eq.~(\ref{transMErpa}) and write the contribution from
a single frequency to the RPA energy-weighted
sum rule as
\begin{equation}
 S_1^{RPA}(\mu)=\Omega_\mu \sum_M | f^M \cdot X^\mu + (-1)^M f^{-M} \cdot Y^\mu | ^2.
\label{ewsr}
\end{equation}
Even if $\mu$ is not a zero mode, one is free to transform to
collective coordinates and momenta \cite{ring},
\begin{eqnarray}
X^\mu = \sqrt{\frac{M_\mu \Omega_\mu}{2}} Q_\mu -i
\sqrt{\frac{1}{2M_\mu\Omega_\mu}} P_\mu \\
Y^\mu = -\sqrt{\frac{M_\mu\Omega_\mu}{2}} Q_{\mu}^* +i
\sqrt{\frac{1}{2M_\mu\Omega_\mu}} P_{\mu}^*.
\end{eqnarray}
Inserting into (\ref{ewsr}) and letting $\Omega_\mu \rightarrow 0$,
there is a finite remainder exactly equal to the rightmost term
of (\ref{eq:corrCom}).  It is of course surprising to find contribution
to the energy-weighted sum rule from `zero excitation energy'. But we remind
the reader that the RPA zero modes are \textit{continuous} states arising from
broken symmetries in the mean-field solution, while the discrete states
are quantized harmonic oscillations about the HF Slater determinant. Part of the 
challenge of RPA is that continuous
modes clearly cannot be treated on the same footing as the discrete states.

We find numerically that the discrepancy in Table \ref{table:s1} is
given exactly by the last term in Eq.~(\ref{eq:corrCom}). Our
\textit{interpretation} of Table \ref{table:s1} and Eq.~(\ref{eq:corrCom})
is missing strength that goes into g.s.-to-g.s. transitions, due
to incomplete symmetry restoration.  Undoubtedly more work remains,
but we hope our results act to inspire further careful investigation.
For example,  for some transition
operators there is no or very small contribution from the zero modes even for
nuclides with deformed HF states; this seems to be associated 
with transitions with high-lying giant resonances, again consistent 
with our interpretation of incomplete restoration of symmetries 
and \textit{low-lying} strength being subsumed into the RPA ground state.

Finally, we wish to discuss the general rule
\begin{equation}
\sum_\nu (E_\nu-E_\mu)
|\langle \mu |F|\nu\rangle|^2 =
\frac{1}{2}\langle \mu|[F,[H,F]]| \mu \rangle.
\label{ewsr1}
\end{equation}
Eq.~(\ref{ewsr1}) is true for any true eigenstate  $|\mu\rangle$, 
and so holds for full shell-model calculations. But 
the HF state is not an eigenstate, and so 
one cannot use the Lanczos moment method 
described in Section \ref{ssec:SM}. 
Instead, we take the HF state and project onto a vector in the basis 
of shell-model states (because shell-model basis states have 
good $J_z$, the projection can only be done easily  for 
even-even nuclides); both $H$ and $F_{KM}$ are matrices in 
the restricted model space and we compute directly 
$(HF-FH)|\mathrm{HF} \rangle$ and dot that vector onto 
$F|\mathrm{HF} \rangle$. One must sum over all shell-model 
states with intermediate values of $J_z$, a tedious but necessary task 
for computing $S_1^{HF}$.

\begin{acknowledgments}
The U.S.~Department of Energy supported this investigation through
grant DE-FG02-96ER40985.
\end{acknowledgments}

\end{document}